*Research Article*

# Nonequilibrium Quantum Systems: Divergence between Global and Local Descriptions

**Pedro D. Manrique,[1] Ferney Rodríguez,[2] Luis Quiroga,[2] and Neil F. Johnson[1]**

[1]*Physics Department, University of Miami, Coral Gables, FL 33126, USA*
[2]*Departamento de Física, Universidad de Los Andes, A.A. 4976, Bogotá, Colombia*

Correspondence should be addressed to Neil F. Johnson; njohnson@physics.miami.edu

Even photosynthesis—the most basic natural phenomenon underlying life on Earth—involves the nontrivial processing of excitations at the pico- and femtosecond scales during light-harvesting. The desire to understand such natural phenomena, as well as interpret the output from ultrafast experimental probes, creates an urgent need for accurate quantitative theories of open quantum systems. However it is unclear how best to generalize the well-established assumptions of an isolated system, particularly under nonequilibrium conditions. Here we compare two popular approaches: a description in terms of a direct product of the states of each individual system (i.e., a *local* approach) versus the use of new states resulting from diagonalizing the whole Hamiltonian (i.e., a *global* approach). The main difference lies in finding suitable operators to derive the Lindbladian and hence the master equation. We show that their equivalence fails when the system is open, in particular under the experimentally ubiquitous condition of a temperature gradient. By solving for the steady state populations and calculating the heat flux as a test observable, we uncover stark differences between the formulations. This divergence highlights the need to establish rigorous ranges of applicability for such methods in modeling nanoscale transfer phenomena—including during the light-harvesting process in photosynthesis.

## 1. Introduction

The rapid recent development of quantum control, quantum engineering, and information processing at the nanoscale in biological, chemical, and condensed matter systems has led to a crucial need to improve our understanding of *open* quantum systems. The typical physics assumptions of an isolated system, particularly under nonequilibrium conditions, cannot hold for systems probed on the quantum scale at optical pico- and femtosecond scales [1]. The accurate description of excitation dynamics at these ultrafast scales is essential to understand fundamental processes for Life on Earth such as photosynthesis [2–4]. For such reasons, theoretical physicists have begun to develop theoretical and experimental tools to study the dynamical behavior of an open system interacting with its environment. The key lies in identifying an accurate way of removing the environmental degrees of freedom and hence obtain a closed equation of motion for the reduced system of interest [5–9]. Quantum master equations have been tested for exactly solvable interacting harmonic systems in thermal equilibrium and nonequilibrium conditions providing reliable results [10]. However, many systems of interest (e.g., optically probed semiconductor quantum dots or biological photosynthetic processes) show fluctuations which are far from equilibrium. Moreover, recent evidence suggests that excitation energy transfer in biological systems, particularly photosynthetic membranes, might involve some level of quantum coherence [11, 12].

In this paper, we provide analytic results for the steady state of an open quantum system interacting with two reservoirs at different temperatures following local and global approximations. This mimics the situation in photosynthetic membrane reaction centers, and elsewhere in natural and artificial systems, in which transfer occurs between few-level molecular complexes that in turn may be coupled to reservoirs with different effective temperatures. In both cases we use the same assumptions and the same procedure; the only difference is the election of the free Hamiltonian and, therefore, the basis set. The local approach is commonly used to model incoherent transfer phenomena in small quantum systems [4], while the global approach is a more rigorous way to calculate the quantum properties such as coherence and entanglement [13, 14]. For the simple case of an interacting qubit dimer where one of the qubits is coupled with a thermal bath, the approximations are equivalent only for the case of zero bath temperature [15]. Similar studies have been performed for two interacting quantum harmonic oscillators under nonequilibrium thermal conditions, where the local approach is found to violate the second law of thermodynamics for the nonsymmetrical case [16] (nonidentical systems). This discrepancy represents a serious challenge for modeling of quantum systems. We consider an open system comprising interacting subunits (qubits), which could provide a simple representation of interacting two-level systems in a photosynthetic membrane reaction center. The question of how to solve the problem can then be addressed in two ways. (i) Diagonalize the Hamiltonian of the open system and solve the problem in terms of the diagonal basis set (i.e., *global* approach). (ii) Use the direct product of the individual basis set of the interacting subunits (i.e., *local* approach). We here apply both these methods in parallel and show explicitly how each formulation leads to a different result when the number of interacting subunits is greater than one.

This paper is divided in four parts. In the second part the methods are presented for an arbitrary system. Analytic expressions for populations and heat current are derived and applied to a two-level system. In the third part, the quantum system is extended to an interacting qubit dimer where the quantities are calculated. The fourth part is devoted to analysis and conclusions.

## 2. Formalism

Consider a quantum system under nonequilibrium thermal conditions. Each reservoir is modeled as an infinite collection of harmonic oscillators in thermal equilibrium at temperature given by $\beta_j = 1/k_B T_j$, $j = 1, 2$. We assume that the coupling strength between the reservoirs and the central subsystem is weak; hence we can express the total density operator as a direct product of the reduced density operators of the open system $\widehat{\rho}_s$ and the reservoirs $\widehat{\rho}_1$ and $\widehat{\rho}_2$: $\widehat{\rho} = \widehat{\rho}_s \otimes \widehat{\rho}_1 \otimes \widehat{\rho}_2$. The Hamiltonian of the whole system is then

$$\widehat{H} = \widehat{Q} + \widehat{R}_1 + \widehat{R}_2 + \widehat{S}_1 + \widehat{S}_2, \tag{1}$$

where $\widehat{Q}$ is the Hamiltonian of the open system and the terms $\widehat{R}_j$ and $\widehat{S}_j$ are the reservoir Hamiltonian and the interaction term associated with the reservoir $j$, respectively. The usual route to this problem is to solve directly the second-order expansion of the Liouville von Neumann equation. This is achieved by obtaining the reduced density operator for the central subsystem through a partial trace over the reservoirs degrees of freedom. Within the interaction picture representation, the dynamics for the whole system is given by

$$\frac{d\widehat{\rho}}{dt} = -i\left[\widehat{H}_I(t), \widehat{\rho}(0)\right] - \int_0^t dt_1 \left[\widehat{H}_I(t), \left[\widehat{H}_I(t_1), \widehat{\rho}(t_1)\right]\right]. \tag{2}$$

The Hamiltonian in the interaction picture representation is defined as $\widehat{H}_I(t) = \widehat{U}_0^\dagger \widehat{H}_I \widehat{U}_0$, where $\widehat{U}_0$ is the evolution unitary operator of the free system. The free system is considered as a subsystem whose solutions are well known. For the present work, we use the two methods to describe the free system: global and local.

### 2.1. Global Approach.
For this global approach, it is useful to express the interaction term in the form

$$\widehat{S}_j = \sum_\mu \widehat{V}_{j,\mu} \widehat{f}_{j,\mu} = \sum_\mu \widehat{V}_{j,\mu}^\dagger \widehat{f}_{j,\mu}^\dagger, \tag{3}$$

where the operators $\widehat{V}_{j,\mu}$ act over the open system's degrees of freedom while the operators $\widehat{f}_{j,\mu}$ act on the reservoir $j$. The operators $\widehat{V}_{j,\mu}$ are chosen in such a way that they follow the following commutation relationship:

$$\left[\widehat{Q}, \widehat{V}_{j,\mu}\right] = \omega_\mu \widehat{V}_{j,\mu}. \tag{4}$$

This decomposition is always possible [5, 10, 17]. For instance, for an operator $\widehat{V}_\mu = |i\rangle\langle i|\widehat{S}|k\rangle\langle k|$, where $|i\rangle$ and $|k\rangle$ are eigenstates of $\widehat{Q}$, it results in $\omega_\mu = \langle i|\widehat{Q}|i\rangle - \langle k|\widehat{Q}|k\rangle$. The dynamics, as was mentioned above, is governed by the Liouville von Neumann equation of motion $d\widehat{\rho}/dt = -i[\widehat{H}, \widehat{\rho}]$. By using the Born-Markov approximation, the dynamics of the open system in terms of its reduced density operator $\widehat{\rho}_s$ is given by

$$\frac{d}{dt}\widehat{\rho}_s = -i\left[\widehat{Q}, \widehat{\rho}_s\right] - \sum_{j=1}^2 \mathscr{L}_j(\widehat{\rho}_s), \tag{5}$$

where $\mathscr{L}_j$ is the Lindblad superoperator associated with reservoir $j$ [13]

$$\mathscr{L}_j(\widehat{\rho}_s) = \sum_{\mu,\nu} J_{\mu,\nu}^{(j)}(\omega_\nu) \left\{ \left[\widehat{V}_{j,\mu}, \left[\widehat{V}_{j,\nu}^\dagger, \widehat{\rho}_s\right]\right] \right. \\ \left. - \left(1 - e^{\beta_j \omega_\nu}\right) \left[\widehat{V}_{j,\mu}, \widehat{V}_{j,\nu}^\dagger \widehat{\rho}_s\right] \right\}, \tag{6}$$

where the indices $\mu$ and $\nu$ run over the complete range of operators and $J_{\mu,\nu}^{(j)}(\omega_\nu)$ is the spectral density is defined by [10, 13, 17]

$$J_{\mu,\nu}^{(j)}(\omega_\nu) = \int_0^\infty d\tau e^{i\omega_\nu \tau} \text{Tr}_{R_j} \left\{ \rho_j \overline{f}_{j,\nu}^\dagger(\tau) \widehat{f}_{j,\mu} \right\}, \tag{7}$$

where $\overline{f}_{j,\nu}(\tau) = e^{-i\widehat{R}_j \tau} \widehat{f}_{j,\nu} e^{i\widehat{R}_j \tau}$ is the interaction picture representation of reservoir operator $\widehat{f}_{j,\nu}$. For each selection of the open system, a new set of operators $\{\widehat{V}_{j,\mu}\}$ is found.

*2.2. Local Approach.* Similarly, the problem can be addressed by using a free Hamiltonian which is formed by summing all zero-point Hamiltonians of each subunit. For a simple subunit such as the qubit, the Hilbert space is spanned by two states with energy gap of $\epsilon$, and the Hamiltonian can be written in terms of the $2 \times 2$ Pauli matrices as $\widehat{Q} = (1/2)\sigma^z \epsilon$. For instance, for the case where the open system is composed by a linear chain of $N$ qubits and considering an $XX$-like interaction between them, the Hamiltonian of the open system can be written as $\widehat{Q} = \widehat{Q}_0 + \widehat{Q}_I$, where $\widehat{Q}_0 = \sum_{q=1}^{N} (1/2)\epsilon_q \sigma_q^z$ is the contribution of each subsystem and $\widehat{Q}_I = \sum_{i=1}^{N-1} K_i \hat{\sigma}_i^+ \hat{\sigma}_{i+1}^-$ + h.c. describes the interqubit interaction. Consequently, the free Hamiltonian $\widehat{H}_0$ and interaction Hamiltonian $\widehat{H}_I$ can be identified as

$$\widehat{H}_0 = \widehat{Q}_0 + \widehat{R}_1 + \widehat{R}_2,$$
$$\widehat{H}_I = \widehat{Q}_I + \widehat{S}_1 + \widehat{S}_2. \tag{8}$$

Hence in the interaction picture representation, the Hamiltonian can be written as

$$\overline{H}_I(t) = \overline{Q}_I(t) + \overline{S}_1(t) + \overline{S}_2(t),$$
$$\overline{Q}_I(t) = \sum_{i=1}^{N-1} K_i \hat{\sigma}_i^+ \hat{\sigma}_{i+1}^- e^{i(\epsilon_i - \epsilon_{i+1})t} + \text{h.c.}, \tag{9}$$
$$\overline{S}_j(t) = \sum_k g_k^{(j)} \hat{a}_{k,j} \sigma_\lambda^+ e^{i(\epsilon_\lambda - \epsilon_{j,k})t} + \text{h.c.},$$

where $\hat{a}_{k,j}^\dagger$ creates an excitation in mode $k$ of reservoir $j$ with a coupling strength of $g_k^{(j)}$ and energy $\epsilon_{j,k}$. The subindex $\lambda$ labels the subunit interacting with the reservoir $j$; that is, $\lambda = 1$ when $j = 1$ and $\lambda = N$ for $j = 2$. Note that for the energy associated with the central system we use only one subindex, for example, $\epsilon_\lambda$, while for the energy associated with one mode of the bath we use two, for example, $\epsilon_{j,k}$. We use the commutation relations for Pauli operators ($[\sigma^+, \sigma^-] = \sigma^z$ and $[\sigma^z, \sigma^\pm] = \pm 2\sigma^\pm$) and for bosonic operators ($[\hat{a}_k, \hat{a}_{k'}^\dagger] = \delta_{k,k'}$). As a result of this transformation, we can use (2) to solve the problem. Specifically, we take the partial trace over the reservoirs and use the Born-Markov approximation. Furthermore, we take the continuous limit for the reservoir frequencies and the wide band limit on the interaction with the central subsystem. This procedure yields to a delta function in the energies ($\delta(\epsilon_\lambda - \epsilon_{j,k})$) that collapses all the energy spectra of reservoir $j$ into the energy of the subunit $\lambda$. Under these conditions, the quantum optical master equation for a chain of qubits, whose endpoints ($\lambda = 1$ and $\lambda = N$) interact with bosonic reservoirs, is given by

$$\frac{d\hat{\rho}_c}{dt} = -i\left[\overline{Q}_I(t), \hat{\rho}_c\right]$$
$$+ \sum_{j=1}^{2} J^{(j)}(\epsilon_\lambda) e^{\beta_j \epsilon_\lambda} \left(\hat{\sigma}_\lambda^- \hat{\rho}_c \hat{\sigma}_\lambda^+ - \frac{1}{2}\{\hat{\sigma}_\lambda^+ \hat{\sigma}_\lambda^-, \hat{\rho}_c\}\right) \tag{10}$$
$$+ J^{(j)}(\epsilon_\lambda) \left(\hat{\sigma}_\lambda^+ \hat{\rho}_c \hat{\sigma}_\lambda^- - \frac{1}{2}\{\hat{\sigma}_\lambda^- \hat{\sigma}_\lambda^+, \hat{\rho}_c\}\right),$$

where $J^{(j)}$ denotes the spectral density function associated with the interaction between the qubit $\lambda$ and the reservoir $j$ given in terms of the spontaneous emission rate $\gamma_j$ and the Bose-Einstein distribution $\overline{N}_j(\epsilon) = (e^{\beta_j \epsilon} - 1)^{-1}$

$$J^{(j)}(\epsilon) = \gamma_j \overline{N}_j(\epsilon). \tag{11}$$

For simplicity, we consider $\gamma_j = 1$.

*2.3. Observable: Thermal Energy.* As a test observable, we use the steady state thermal energy transferred from the reservoir to the quantum system given the nonzero thermal bias. The steady state heat flux is defined as the trace of the product of the Liouvillian superoperator with the subsystem Hamiltonian:

$$\mathcal{Q}_j = \text{Tr}\{\widehat{Q}\mathcal{L}_j\}. \tag{12}$$

The resultant expression for the heat flux in the global approach can be written in the following compact form:

$$\mathcal{Q}_j = \sum_{\mu,\nu} \omega_\mu J_{\mu,\nu}^{(j)}(\omega_\nu) \left\{\langle \widehat{V}_{j,\nu}^\dagger \widehat{V}_{j,\mu} \rangle - e^{\beta_j \omega_\nu} \langle \widehat{V}_{j,\mu} \widehat{V}_{j,\nu}^\dagger \rangle\right\}. \tag{13}$$

In a similar way, the result for the heat flux in the local approach is found to be

$$\mathcal{Q}_j = \epsilon_\lambda J^{(j)}(\epsilon_\lambda) \left\{\langle \hat{\sigma}_\lambda^- \hat{\sigma}_\lambda^+ \rangle - e^{\beta_j \epsilon_\lambda} \langle \hat{\sigma}_\lambda^+ \hat{\sigma}_\lambda^- \rangle\right\}. \tag{14}$$

As expected, formula (14) gives a zero flux value when the reservoirs are set at the same temperature. Furthermore, it can be seen that the global and local approaches lead to the same result when applied to a single qubit under nonequilibrium thermal conditions where the system operators are the same, $\widehat{V}_{j,\mu}^{(\dagger)} = \hat{\sigma}^{+(-)}$. The steady state population of the excited state $n$ of one qubit system with energy gap $\epsilon$ is found to be

$$n = \langle \hat{\sigma}^+ \hat{\sigma}^- \rangle = \frac{1}{2} e(\epsilon), \tag{15}$$

where $e_j$ is a simple universal function defined as

$$e(\omega_j) = \frac{\overline{N}_1(\omega_j) + \overline{N}_2(\omega_j)}{1 + \overline{N}_1(\omega_j) + \overline{N}_2(\omega_j)} \leq 1. \tag{16}$$

Furthermore, the heat flux for this system is found to be

$$\mathcal{Q}_1 = \epsilon J^{(1)}(\epsilon) J^{(2)}(\epsilon) \frac{e^{\beta_2 \epsilon} - e^{\beta_1 \epsilon}}{J^{(1)}(\epsilon)(1 + e^{\beta_1 \epsilon}) + J^{(2)}(\epsilon)(1 + e^{\beta_2 \epsilon})}, \tag{17}$$

$$= \frac{\epsilon}{2}(1 - e(\epsilon))(\overline{N}_1(\epsilon) - \overline{N}_2(\epsilon)). \tag{18}$$

As can be seen, formula (18) leads to a zero value when the thermal bias is set as zero. In addition, note the heat flux is always positive for positive bias ($\overline{N}_1 > \overline{N}_2$) and negative, otherwise.

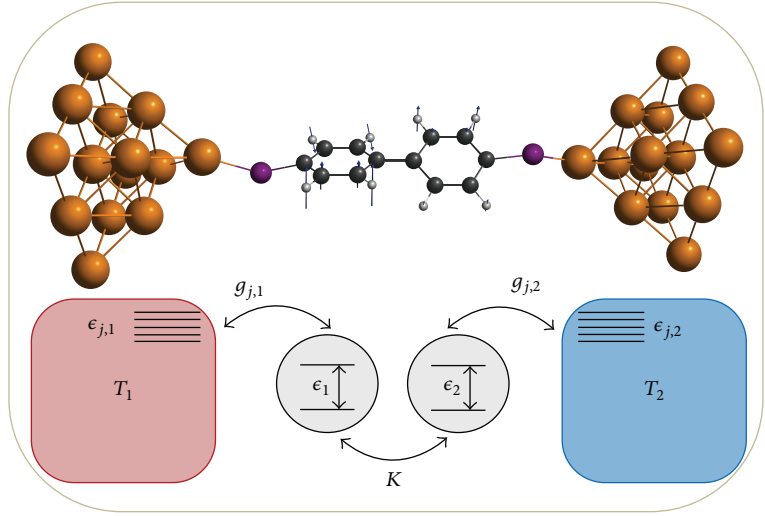

Figure 1: Extended quantum system interacting with two thermal reservoirs.

## 3. Divergence for a Dimer

We consider a dimer composed of two interacting qubits, where each quit is connected to a different reservoir in thermal equilibrium at temperature $T_j$ ($j = 1, 2$). Figure 1 shows schematically the system to be studied. The Hamiltonian of the qubit dimer is

$$\widehat{Q} = \frac{1}{2}\epsilon_1 \widehat{\sigma}_1^z + \frac{1}{2}\epsilon_2 \widehat{\sigma}_2^z + K\left(\widehat{\sigma}_1^+ \widehat{\sigma}_2^- + \widehat{\sigma}_1^- \widehat{\sigma}_2^+\right), \quad (19)$$

where $\widehat{\sigma}_1 = \widehat{\sigma} \otimes I$ and $\widehat{\sigma}_2 = I \otimes \widehat{\sigma}$, with $I$ as the $2 \times 2$ identity matrix. In addition, we assume that the qubit labeled as $j$ interacts with the bath labeled as $j$ only, for $j = 1, 2$. Hence the interaction Hamiltonian between the dimer and the reservoirs can be written as

$$\widehat{S}_j = \sum_k g_{k,j} \sigma_j^+ \widehat{a}_{k,j} + \text{h.c.} \quad (20)$$

The operators $\widehat{\sigma}_j^\pm$ do not commute with $\widehat{Q}$, so for the global approach it is necessary to find a transformation that allows us to write the interaction Hamiltonian in the form of (3), so that condition (4) is fulfilled. The eigenstates and eigenenergies of $\widehat{Q}$ are $|s_1\rangle = |0, 0\rangle$ ($E_1 = -((\epsilon_1 + \epsilon_2)/2)$), $|s_2\rangle = |1, 1\rangle$ ($E_2 = (\epsilon_1 + \epsilon_2)/2$), $|s_3\rangle = c_{3,1}|1, 0\rangle + c_{3,2}|0, 1\rangle$ ($E_3 = \alpha$), and $|s_4\rangle = c_{4,1}|1, 0\rangle + c_{4,2}|0, 1\rangle$ ($E_4 = -\alpha$), where the amplitudes and constants are given by

$$c_{3,1} = \frac{K}{\sqrt{2\alpha^2 - \alpha \Delta\epsilon}}, \quad c_{3,2} = \frac{\alpha - \Delta\epsilon/2}{\sqrt{2\alpha^2 - \alpha \Delta\epsilon}},$$

$$c_{4,1} = \frac{K}{\sqrt{2\alpha^2 + \alpha \Delta\epsilon}}, \quad c_{4,2} = -\frac{\alpha + \Delta\epsilon/2}{\sqrt{2\alpha^2 + \alpha \Delta\epsilon}}, \quad (21)$$

$$\alpha = \sqrt{K^2 + \frac{\Delta\epsilon^2}{4}}, \quad \Delta\epsilon = \epsilon_1 - \epsilon_2.$$

The transformation of the coupling operators from the individual qubits basis set into the dimer diagonal basis set can be calculated as

$$\widehat{\sigma}_j = \sum_{p=1}^{4} \sum_{q=1}^{4} |s_p\rangle \langle s_p|\widehat{\sigma}_j|s_q\rangle \langle s_q|. \quad (22)$$

With this transformation, condition (4) is fulfilled $[\widehat{Q}, |s_p\rangle\langle s_q|] = (E_p - E_q)|s_p\rangle\langle s_q|$. In this way, the operators can be found to be

$$\widehat{V}_{j,1} = \left[c_{3,2}\delta_{j,1} + c_{3,1}\delta_{j,2}\right]|s_2\rangle\langle s_3|, \quad \omega_1 = E_2 - E_3,$$

$$\widehat{V}_{j,2} = \left[c_{4,2}\delta_{j,1} + c_{4,1}\delta_{j,2}\right]|s_2\rangle\langle s_4|, \quad \omega_2 = E_2 - E_4,$$

$$\widehat{V}_{j,3} = \left[c_{3,1}\delta_{j,1} + c_{3,2}\delta_{j,2}\right]|s_3\rangle\langle s_1|, \quad \omega_3 = E_3 - E_1,$$

$$\widehat{V}_{j,4} = \left[c_{4,1}\delta_{j,1} + c_{4,2}\delta_{j,2}\right]|s_4\rangle\langle s_1|, \quad \omega_4 = E_4 - E_1. \quad (23)$$

For simplicity we consider the symmetric case where the energy gap of each qubit is the same: $\epsilon_1 = \epsilon_2 = \epsilon$. In particular, we have that $\omega_1 = \omega_4 = |\epsilon - K|$, $\omega_2 = \omega_3 = \epsilon + K$, and the amplitudes are $c_{3,1} = c_{3,2} = c_{4,1} = 1/\sqrt{2}$ and $c_{4,2} = -1/\sqrt{2}$. The steady state populations can be calculated for each of the qubits:

$$n_1 = \langle \widehat{\sigma}_1^+ \widehat{\sigma}_1^- \rangle = \rho_{22}^{(G)} + \frac{1}{2}\left(\rho_{33}^{(G)} + \rho_{44}^{(G)} + 2\,\text{Re}\left\{\rho_{34}^{(G)}\right\}\right),$$

$$n_2 = \langle \widehat{\sigma}_2^+ \widehat{\sigma}_2^- \rangle = \rho_{22}^{(G)} + \frac{1}{2}\left(\rho_{33}^{(G)} + \rho_{44}^{(G)} - 2\,\text{Re}\left\{\rho_{34}^{(G)}\right\}\right), \quad (24)$$

where the matrix elements $\rho_{ij}^{(G)} = \langle s_i|\widehat{\rho}_s|s_j\rangle$ are calculated in the diagonal basis. For the weak-coupling case ($\epsilon > K$), the populations are found to be ($K = 1$)

$$\rho_{11}^{(G)} = \left(1 - \frac{e_1}{2}\right)\left(1 - \frac{e_2}{2}\right), \quad \rho_{22}^{(G)} = \frac{e_1}{2}\frac{e_2}{2},$$

$$\rho_{33}^{(G)} = \left(1 - \frac{e_1}{2}\right)\frac{e_2}{2}, \quad \rho_{44}^{(G)} = \frac{e_1}{2}\left(1 - \frac{e_2}{2}\right), \quad (25)$$

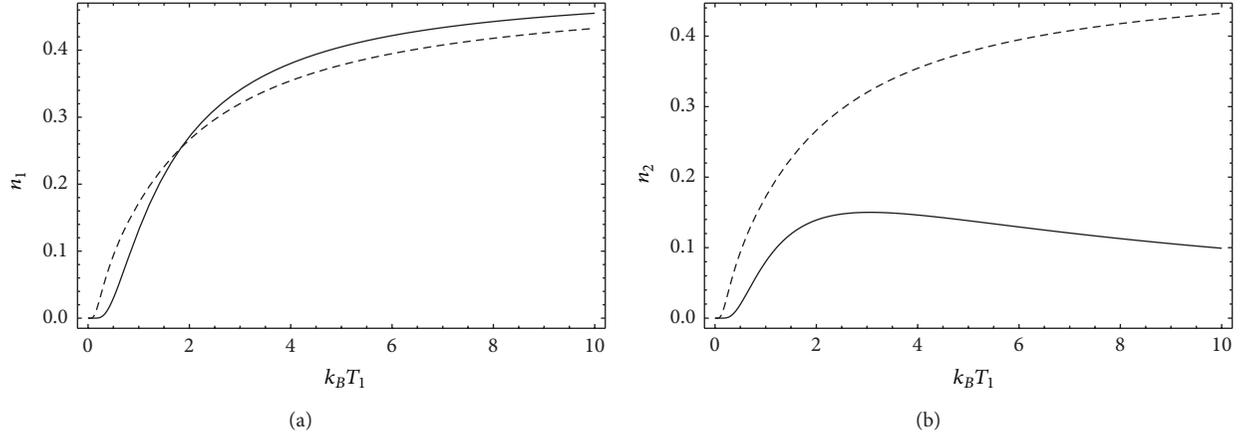

FIGURE 2: Steady state populations $n_1$ (a) and $n_2$ (b) as a function of the temperature of reservoir 1 for both approximations. The solid lines represent the local approach while the dashed lines represent the global approach. In all cases, the temperature of reservoir 2 is set as zero, $K = 1$ and $\epsilon = 1.5$.

where $e_j = e(\omega_j)$. For the steady state, the matrix element $\rho_{34}^{(G)}$ is zero; therefore the populations take the simple form of $n_{1,2} = (e_1 + e_2)/4$. In a similar way the populations for the local approach can be derived by solving (10) for $N = 2$ leading to $n_{1(2)} = \rho_{11}^{(L)} + \rho_{22(33)}^{(L)}$, where the matrix elements $\rho_{ij}^{(L)}$ are calculated in the local basis; that is, $\rho_{11}^{(L)} = \langle 11|\hat{\rho}_s|11\rangle$, $\rho_{22}^{(L)} = \langle 10|\hat{\rho}_s|10\rangle$, and $\rho_{33}^{(L)} = \langle 01|\hat{\rho}_s|01\rangle$. The following results follow:

$$n_1 = \frac{2K^2 e(\epsilon) + \left(1 + 2\overline{N}_2(\epsilon)\right)\overline{N}_1(\epsilon)}{4K^2 + \left(1 + 2\overline{N}_1(\epsilon)\right)\left(1 + 2\overline{N}_2(\epsilon)\right)},$$

$$n_2 = \frac{2K^2 e(\epsilon) + \left(1 + 2\overline{N}_1(\epsilon)\right)\overline{N}_2(\epsilon)}{4K^2 + \left(1 + 2\overline{N}_1(\epsilon)\right)\left(1 + 2\overline{N}_2(\epsilon)\right)}. \quad (26)$$

In Figure 2 we can see how the populations change as the temperature of reservoir 1 changes while the temperature of reservoir 2 is set to be zero. The two approaches for $n_1$ converge when $k_B T_1 \gg \epsilon$. On the other hand, population $n_2$ decays in the local approach as $k_B T_1 \gg \epsilon$ while the global approach predicts an asymptotic growth to the mixed state of $1/2$. This divergence in predictions suggests that one of the approaches is not correct.

Another way to see this discrepancy is by looking at the heat flux. Using the system operators (23), the steady state heat flux (13) for reservoir 1 is

$$\mathcal{Q}_1 = \sum_{i=1}^{2} \omega_i J^{(1)}(\omega_i) J^{(2)}(\omega_i)$$
$$\cdot \frac{e^{\beta_2 \omega_i} - e^{\beta_1 \omega_i}}{\left(1 + e^{\beta_1 \omega_i}\right) J^{(1)}(\omega_i) + \left(1 + e^{\beta_2 \omega_i}\right) J^{(2)}(\omega_i)} \quad (27)$$
$$= \frac{\omega_1}{2}\left(1 - e_1\right)\left(\overline{N}_1(\omega_1) - \overline{N}_2(\omega_1)\right) - (1 \longleftrightarrow 2).$$

The heat flux is expressed as a sum over the energy channels $\omega_1$ and $\omega_2$ that depend on the interqubit coupling $K$ and qubit energy gap $\epsilon$. The flux is always positive for positive thermal bias. On the other hand, the steady state heat flux from reservoir 1 to qubit 1, calculated in the local approach for the symmetric case, can be found to be

$$\mathcal{Q}_1 = \frac{\epsilon J^{(1)}(\epsilon) J^{(2)}(\epsilon)\left(e^{\beta_2 \epsilon} - e^{\beta_1 \epsilon}\right)}{J^{(1)}(\epsilon)\left(1 + e^{\beta_1 \epsilon}\right) + J^{(2)}(\epsilon)\left(1 + e^{\beta_2 \epsilon}\right)}$$
$$\cdot \frac{4K^2}{4K^2 + J^{(1)}(\epsilon) J^{(2)}(\epsilon)\left(1 + e^{\beta_1 \epsilon}\right)\left(1 + e^{\beta_2 \epsilon}\right)} \quad (28)$$
$$= \left[\frac{\epsilon}{2}\left(1 - e(\epsilon)\right)\left(\overline{N}_1(\epsilon) - \overline{N}_2(\epsilon)\right)\right]$$
$$\cdot \frac{4K^2}{4K^2 + J^{(1)}(\epsilon) J^{(2)}(\epsilon)\left(1 + e^{\beta_1 \epsilon}\right)\left(1 + e^{\beta_2 \epsilon}\right)}.$$

The local approach for the dimer leads to an expression for the heat flux expression equal to the one for the monomer weighted by a positive function that depends on the interqubit coupling $K$ and the reservoirs' temperature. This function ensures that the flux tends to zero as the interqubit coupling decreases and remains finite for large $K$. This is reasonable since the qubits are weakly coupled (i.e., $K \to 0$), and therefore the heat transferred should decrease. However it also shapes the flux in such a way that an optimal value is found; that is, the heat flux exhibits a maximum with the temperature and then decays to zero as the thermal bias grows. This behavior is not found for either the monomer or the dimer in the global approach.

As an illustration, Figure 3 shows the observable $\mathcal{Q}_1$ for both approximations and three different values of the qubit energy gap $\epsilon$, as a function of the temperature difference. To simplify the presentation, we have set the temperature of the second reservoir to be close to zero and have depicted the flux as a function of $k_B T_1$. We can clearly see the maximum of $\mathcal{Q}_1$ for a specific value of the energy $k_B T_1$ in the local approach, while the result for the global approach grows asymptotically to $\epsilon$ as the thermal bias increases.

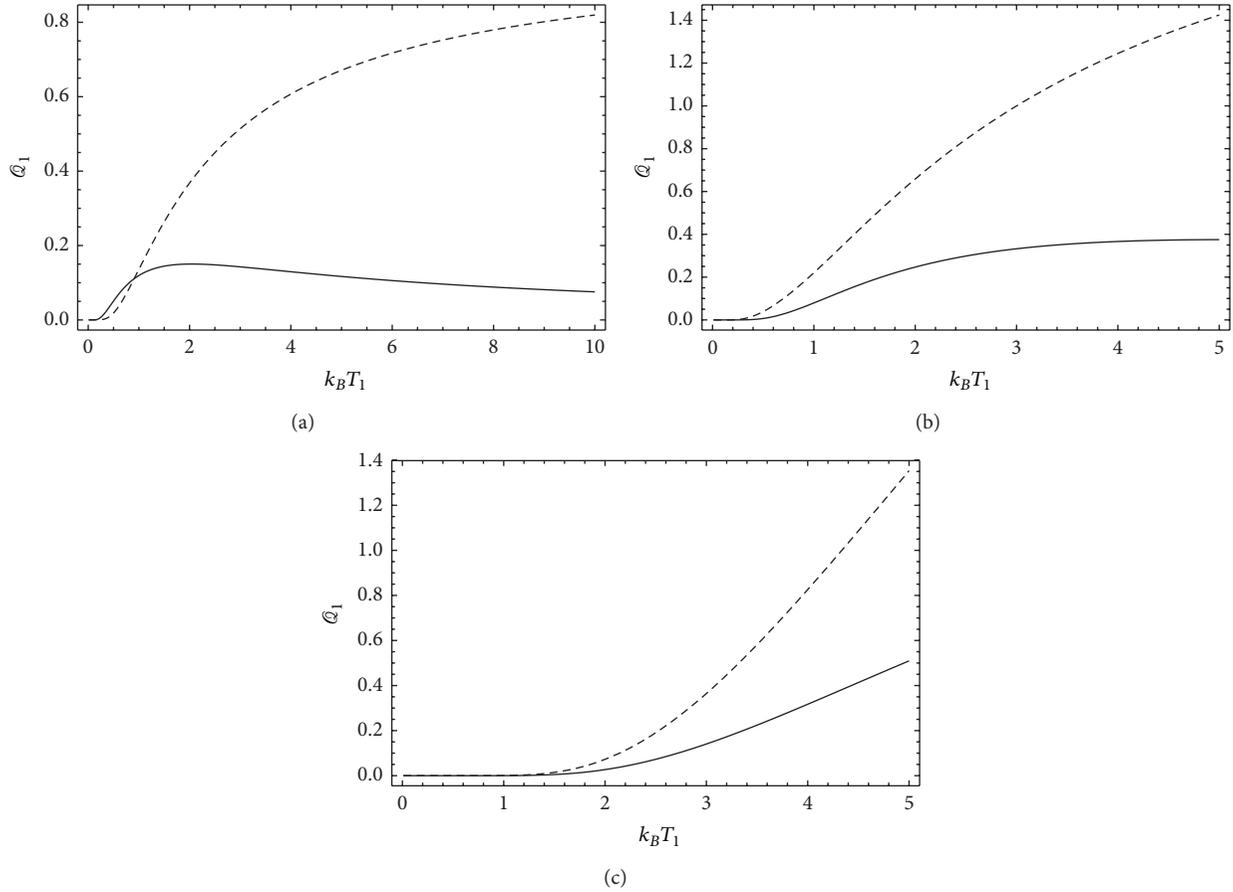

FIGURE 3: Dimer heat flux $\mathcal{Q}_1$ as a function of the temperature of reservoir 1 in both approximations. The solid lines represent the local approach while the dashed lines represent the global approach. The steady state heat flux is calculated for three different values of the qubit energy gap $\epsilon$: (a) $\epsilon = 1.001K$, (b) $\epsilon = 2.5K$, and (c) $\epsilon = 10K$. In all the cases we have set the temperature of reservoir 2 as zero and $K = 1$.

The dimer system thus provides the simplest nontrivial physical scenario where a quantitative comparison of quantum nonequilibrium thermal quantities can be performed: those obtained within a rigorous global approach and results coming from a less rigorous and hence restricted in validity, local approach. Furthermore, our results as tested in simple systems show that some physical magnitudes calculated within the local framework could be misleading. This work suggests various future avenues of research, one of which concerns the systematic analysis of the scaling of the divergence "distance" between different approach results as the system size (complexity) and the separation from equilibrium increase.

## 4. Conclusions

We have shown that even in the symmetrical case, the problem of an open quantum system interacting with two reservoirs at different temperatures leads, within local and global approximations, to different results when the number of interacting subunits in the open system is greater than one. The formulations are equivalent and identical for the monomer. By contrast, the results for an interacting dimer open several urgent questions about the range of applicability of the underlying approximations and hence methods. In the global approach, the populations for both qubits are identical in the steady state. On the contrary, the local approach predicts that the population of the qubit interacting with the cold reservoir is always smaller than the population of the other qubit. Second, the local approach predicts a maximum in the heat flux as a function of the temperature gradient, followed by a gradual decay to zero as the thermal bias grows. By contrast, the global approach predicts a saturation of the flux as the bias increases. Finally, we note that the outcome of the local approach in the strong interqubit coupling limit concludes that the dimer can be modeled as a single qubit which resembles the properties of a classical system. Future work with larger numbers of qubits will elucidate the differences in these approaches, as will a careful comparison to future experiments which are able to distinguish between the divergences in their predictions.

## Conflict of Interests

The authors declare that there is no conflict of interests regarding the publication of this paper.

## References

[1] G. M. Moy, J. J. Hope, and C. M. Savage, "Born and Markov approximations for atom lasers," *Physical Review A—Atomic, Molecular, and Optical Physics*, vol. 59, no. 1, pp. 667–675, 1999.

[2] A. Olaya-Castro, C. F. Lee, F. F. Olsen, and N. F. Johnson, "Efficiency of energy transfer in a light-harvesting system under quantum coherence," *Physical Review B*, vol. 78, no. 8, Article ID 085115, 2008.

[3] F. Caycedo-Soler, F. J. Rodríguez, L. Quiroga, and N. F. Johnson, "Light-harvesting mechanism of bacteria exploits a critical interplay between the dynamics of transport and trapping," *Physical Review Letters*, vol. 104, no. 15, Article ID 158302, 2010.

[4] L. Campos Venuti and P. Zanardi, "Excitation transfer through open quantum networks: a few basic mechanisms," *Physical Review B*, vol. 84, no. 13, Article ID 134206, 11 pages, 2011.

[5] H.-P. Breuer and F. Petruccione, *The Theory of Open Quantum Systems*, Oxford University Press, 2006.

[6] A. J. Leggett, S. Chakravarty, A. T. Dorsey, M. P. A. Fisher, A. Garg, and W. Zwerger, "Dynamics of the dissipative two-state system," *Reviews of Modern Physics*, vol. 59, no. 1, pp. 1–85, 1987.

[7] H.-P. Breuer, "Non-Markovian quantum dynamics and the method of correlated projection super-operators," in *Theoretical Foundations of Quantum Information Processing and Communication*, vol. 787 of *Lecture Notes in Physics*, pp. 125–139, Springer, Berlin, Germany, 2010.

[8] H.-P. Breuer, "Non-Markovian generalization of the Lindblad theory of open quantum systems," *Physical Review A*, vol. 75, no. 2, 2007.

[9] U. Weiss, *Quantum Dissivative Systems*, World Scientific, Singapore, 1999.

[10] Á. Rivas, A. D. K. Plato, S. F. Huelga, and M. B. Plenio, "Markovian master equations: a critical study," *New Journal of Physics*, vol. 12, Article ID 113032, 2010.

[11] Y.-C. Cheng and G. R. Fleming, "Dynamics of light harvesting in photosynthesis," *Annual Review of Physical Chemistry*, vol. 60, pp. 241–262, 2009.

[12] A. W. Chin, J. Prior, R. Rosenbach, F. Caycedo-Soler, S. F. Huelga, and M. B. Plenio, "The role of non-equilibrium vibrational structures in electronic coherence and recoherence in pigment-protein complexes," *Nature Physics*, vol. 9, no. 2, pp. 113–118, 2013.

[13] L. Quiroga, F. J. Rodríguez, M. E. Ramírez, and R. París, "Nonequilibrium thermal entanglement," *Physical Review A*, vol. 75, Article ID 032308, 2007.

[14] J. C. Castillo, F. J. Rodríguez, and L. Quiroga, "Enhanced violation of a Leggett-Garg inequality under nonequilibrium thermal conditions," *Physical Review A: Atomic, Molecular, and Optical Physics*, vol. 88, no. 2, Article ID 022104, 2013.

[15] G. L. Deçordi and A. Vidiella-Barranco, "Coherence and entanglement in a two-qubit system coupled to a finite remperature reservoir: a comparative study," http://arxiv-web3.library.cornell.edu/abs/1406.0528v1.

[16] A. Levy and R. Kosloff, "The local approach to quantum transport may violate the second law of thermodynamics," *Europhysics Letters*, vol. 107, no. 2, Article ID 20004, 2014.

[17] M. Goldman, "Formal theory of spin-lattice relaxation," *Journal of Magnetic Resonance*, vol. 149, no. 2, pp. 160–187, 2001.

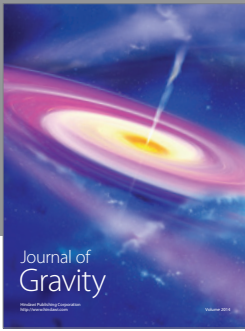 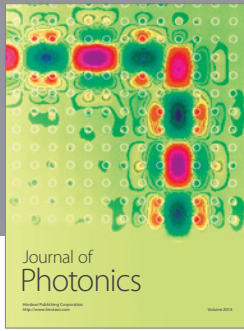 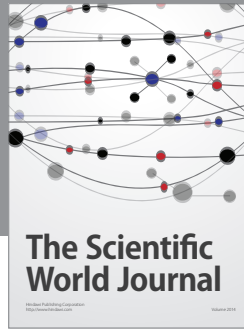 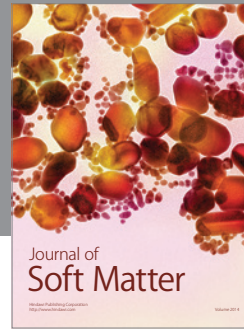 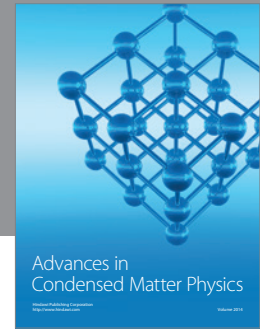
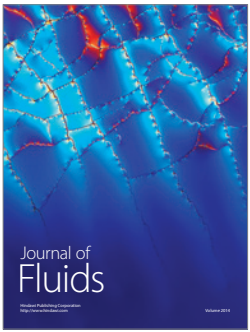 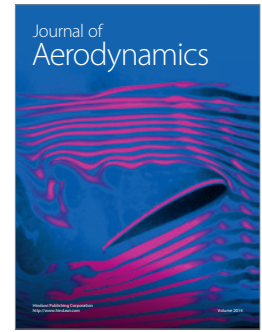
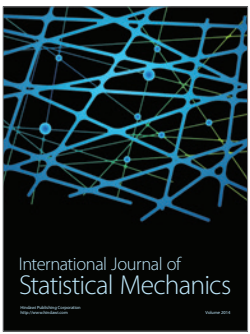 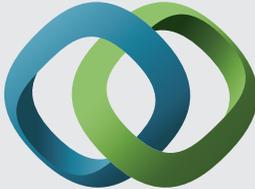 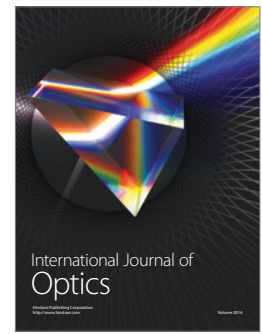
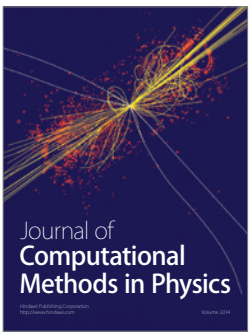 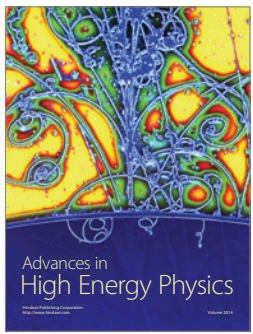 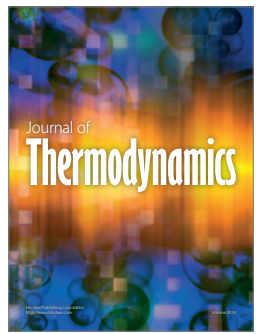 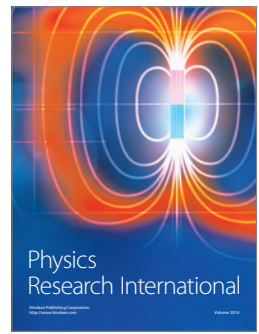 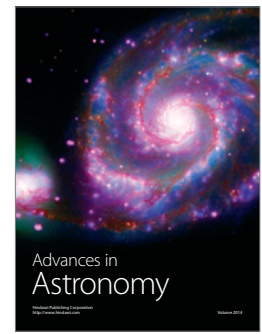
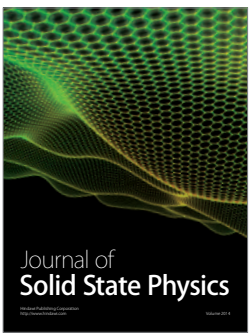 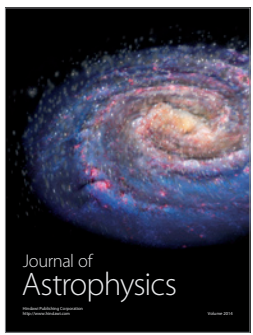 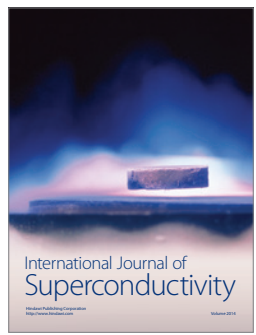 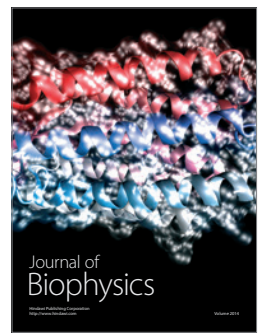 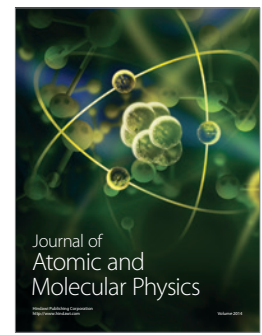

Submit your manuscripts at
http://www.hindawi.com